\PassOptionsToPackage{table,svgnames,dvipsnames}{xcolor}

\def\target{CCR}
\def\IMC{IMC}
\def\CCR{CCR}
\def\WWW{WWW}
\def\SIGCOMM{SIGCOMM}
\def\CONEXT{CONEXT}

\ifx\target\CCR
\documentclass[sigconf,screen=true]{acmart}

\setcopyright{none}
\renewcommand\footnotetextcopyrightpermission[1]{}

\usepackage{fancyhdr}
\fancypagestyle{plain}{\fancyhf{} \fancyfoot[L]{ACM SIGCOMM Computer Communication Review}\fancyfoot[R]{Volume 52 Issue 2, April 2022}}
\fancypagestyle{firstpagestyle}{\fancyhf{} \fancyfoot[L]{ACM SIGCOMM Computer Communication Review}\fancyfoot[R]{Volume 52 Issue 2, April 2022}}

\fi

\ifx\target\IMC
\documentclass[10pt,sigconf,letterpaper,anonymous,screen=true]{acmart}
\setcopyright{none}
\renewcommand\footnotetextcopyrightpermission[1]{}
\fi

\ifx\target\WWW
\documentclass[sigconf,anonymous,screen=true]{acmart}
\setcopyright{none}
\renewcommand\footnotetextcopyrightpermission[1]{}
\fi

\ifx\target\SIGCOMM
\documentclass[sigconf,screen=true,10pt]{acmart}
\setcopyright{none}
\renewcommand\footnotetextcopyrightpermission[1]{}
\fi

\ifx\target\CONEXT
\documentclass[sigconf,screen=true,10pt]{acmart}
\setcopyright{none}
\renewcommand\footnotetextcopyrightpermission[1]{}
\fi

\usepackage[T1]{fontenc}
\usepackage[utf8]{inputenc}
\usepackage[nolist]{acronym}
\usepackage{url}
\def\UrlOrds{\do\*\do\-\do\~\do\'\do\"\do\-}\makeatletter
\g@addto@macro{\UrlBreaks}{\UrlOrds}
\makeatother
\usepackage{booktabs}
\usepackage{multirow}
\usepackage{color,soul}
\usepackage{xcolor}
\usepackage{balance}
\usepackage{tikz}

\usepackage{tabularx}

\usepackage{blindtext}
\usepackage{mdframed}

\newcommand\copyrighttext{%
	\footnotesize The final publication is available at the ACM DL via \url{https://dl.acm.org/doi/10.1145/3544912.3544918}}
	\newcommand\copyrightnotice{%
	\begin{tikzpicture}[remember picture,overlay]
	\node[anchor=south,yshift=10pt] at (current page.south) {\fbox{\parbox{\dimexpr\textwidth-\fboxsep-\fboxrule\relax}{\copyrighttext}}};
	\end{tikzpicture}%
}

\newcommand{\eg}{e.g., }

\newcommand{\ie}{i.e., }

\newcommand{\takeaway}[1]{\textit{\textbf{Takeaway:}} \textit{#1}}

\begin{acronym}
    \acro{DNS}{Domain Name System}
    \acro{EDNS}{Extended \ac{DNS}}
    \acro{DoUDP}{DNS over UDP}
    \acro{DoTCP}{DNS over TCP}
    \acro{RA}{RIPE Atlas}
    \acro{DoT}{DNS over TLS}
    \acro{DoH}{DNS over HTTPS}
    \acro{DoQ}{DNS over QUIC}
    \acro{DoS}{Denial of Service}
    \acro{QoE}{Quality of Experience}
    \acro{EU}{Europe}
    \acro{NA}{North America}
    \acro{AF}{Africa}
    \acro{SA}{South America}
    \acro{OC}{Oceania}
    \acro{AS}{Asia}
    \acro{PoP}{Point of Presence}
    \acro{UDP}{User Datagram Protocol}
    \acro{TCP}{Transmission Control Protocol}
    \acro{APNIC}{Asia Pacific Network Information Centre}
    \acro{RT}{Response Time}
    \acro{TFO}{TCP Fast Open}
    \acro{FQDN}{Fully Qualified Domain Name}
    \acro{RTT}{Round-Trip Time}
    \acro{ISP}{Internet Service Provider}
    \acro{RCODE}{Response Code}
    \acro{TTL}{Time to Live}
    \acro{CPE}{Customer-premises equipment}
    \acro{IAD}{Integrated access device}
    \acro{WG}{Working Group}
    \acro{IETF}{Internet Engineering Task Force}
    \acro{dnsop}{Domain Name System Operations}
    \acro{RR}{Resource Record}
    \acro{IP}{Internet Protocol}
    \acro{MTU}{Maximum Transmission Unit}
    \acro{EDNS(0)}{Extension Mechanisms for DNS}
\end{acronym}

\acrodefplural{AS}[ASes]{Autonomous Systems}
\acrodefplural{TLD}[TLDs]{Top-Level Domains}
\acrodefplural{TTFB}[TTFBs]{Times to First Byte}
\acrodefplural{PoP}[PoPs]{Points of Presence}
\acrodefplural{FQDN}[FQDNs]{Fully Qualified Domain Names}
\acrodefplural{RCODE}[RCODEs]{Response Codes}
\acrodefplural{RR}[RRs]{Resource Records}
\acrodefplural{RT}[RTs]{Response Times}

\begin{document}

\title[Measuring DNS over TCP in the Era of Increasing DNS Response Sizes]{Measuring DNS over TCP in the Era of Increasing\protect\\DNS Response Sizes: A View from the Edge}

\author{Mike Kosek}
\affiliation{
    \institution{Technical University of Munich}
}
\email{kosek@in.tum.de}

\author{Trinh Viet Doan}
\affiliation{
    \institution{Technical University of Munich}
}
\email{doan@in.tum.de}

\author{Simon Huber}
\affiliation{
    \institution{Technical University of Munich}
}
\email{simon.huber@tum.de}

\author{Vaibhav Bajpai}
\affiliation{
    \institution{CISPA Helmholtz Center for Information Security}
}
\email{bajpai@cispa.de}

\begin{abstract}
    The \ac{DNS} is one of the most crucial parts of the Internet.
    Although the original standard defined the usage of \ac{DoUDP} as well as \ac{DoTCP}, UDP has become the predominant protocol used in the \ac{DNS}.
    With the introduction of new \acp{RR}, the sizes of \ac{DNS} responses have increased considerably.
    Since this can lead to \emph{truncation} or \emph{IP fragmentation}, the fallback to \ac{DoTCP} as required by the standard ensures successful \ac{DNS} responses by overcoming the size limitations of \ac{DoUDP}.
    However, the effects of the usage of \ac{DoTCP} by stub resolvers are not extensively studied to this date.
    We close this gap by presenting a view at \ac{DoTCP} from the Edge, issuing 12.1M \ac{DNS} requests from 2,500 probes toward \emph{Public} as well as \emph{Probe} \ac{DNS} recursive resolvers.
    In our measurement study, we observe that \ac{DoTCP} is generally slower than \ac{DoUDP}, where the relative increase in \emph{Response Time} is less than 37\% for most resolvers.
    While optimizations to \ac{DoTCP} can be leveraged to further reduce the response times, we show that support on \emph{Public} resolvers is still missing, hence leaving room for optimizations in the future.
    Moreover, we also find that \emph{Public} resolvers generally have comparable reliability for \ac{DoTCP} and \ac{DoUDP}.
    However, \emph{Probe} resolvers show a significantly different behavior: \ac{DoTCP} queries targeting \emph{Probe} resolvers fail in 3 out of 4 cases, and, therefore, do not comply with the standard.
    This problem will only aggravate in the future: As \ac{DNS} response sizes will continue to grow, the need for \ac{DoTCP} will solidify.
    \copyrightnotice
    
\end{abstract}



\maketitle

\section{Introduction}
\label{sec:introduction}

The \acf{DNS} is one of the most crucial parts of the Internet, taking part in almost every connection of any service.
The original standard defined the usage of \acf{DoUDP} as well as \acf{DoTCP}~\cite{rfc1034,rfc1035}.
However, UDP has become the predominant protocol used in the \ac{DNS}~\cite{dns.centralization,truncation} due to its latency benefits, given its absence of connection establishment and state handling.

With the introduction of new \acp{RR} such as \emph{AAAA} (IPv6 support)~\cite{rfc3596} or \emph{RRSIG} (DNSSEC)~\cite{rfc4034}, the sizes of \ac{DNS} responses have increased considerably~\cite{nlnet.dnssec,rssac.data.icann}.
The most recent efforts to establish a generic format within the \ac{DNS} to provide clients with information on how to access a service using the \emph{SVCB} (Service Binding) \ac{RR}~\cite{ietf-dnsop-svcb-https}, which also provides the configuration required for \emph{TLS Encrypted Client Hello}~\cite{ietf-tls-esni}, will continue this trend of increasing \ac{DNS} response sizes.

To increase the original \ac{DoUDP} response size limit of 512~bytes \cite{rfc1035}, the \emph{\ac{EDNS(0)}}~\cite{rfc2671,rfc6891} were introduced to allow requests and responses of up to 65,535~bytes.
However, when a \ac{DoUDP} request or response exceeds the limit of either the original 512~bytes or the \emph{EDNS(0)} size signaled, it is marked as \emph{truncated}, which results in fallback to \ac{DoTCP}~\cite{rfc1123,rfc7766}: Due to the connection-oriented nature of TCP, \ac{DoTCP} overcomes the size limitations of \ac{DoUDP} and ensures a successful DNS response.

Several studies investigate the \emph{EDNS(0) Buffer Sizes} used by requests issued from recursive resolvers to authoritative servers, finding that buffer sizes falling short of or exceeding the recommended limits remain the predominant sizes~\cite{dns.centralization,truncation}.
While this poses a risk for \emph{truncation} as well as \emph{IP fragmentation}, the effects of these issues on the DNS are extensively studied~\cite{truncation,dnssec.fragmentation,netalyzr.implications,dns.fragmentation,domain.validation}.
Inherently, \ac{DoT} and \ac{DoH} circumvent these issues by using TCP as the underlaying transport protocol.
However, their adoption by recursive resolvers is still low~\cite{dot.doh,dot.doh.2.,TrinhVietDoan.2021}, and both protocols aid the trend of Internet centralization~\cite{dns.centralization,dns.centralization.viet}.
Hence, the need for \ac{DoTCP} will solidify in the future.
A contemporary study on \ac{DoTCP}~\cite{pam-dotcp} has looked at its support in the wild; the authors find a lack of proper TCP fallback and \ac{DoTCP} adoption in numerous cases, although resolvers should already support \ac{DoTCP} as required by the standard~\cite{rfc7766}.
In general, the effects of \ac{DoTCP} usage by stub resolvers in terms of \emph{Failure Rates} and \emph{Response Times} are not yet extensively studied.

We close this gap by presenting a unique view on DNS over TCP from the Edge, evaluating \emph{Failure Rates} (see~\autoref{sub:failure-rates}) and \emph{Response Times} (see~\autoref{sub:response-times}).
Using the \ac{RA} platform~\cite{ripencc:ipj:2015}, we issue 12.1M \ac{DNS} requests from the stub resolvers of 2,500 probes toward \emph{Public} as well as \emph{Probe} \ac{DNS} recursive resolvers over both \ac{DoTCP} and \ac{DoUDP} (see~\autoref{sec:methodology}).

\textbf{Failure Rates.}
While failure rates over \ac{DoTCP} are comparable with \ac{DoUDP} for \emph{Public} resolvers, \ac{DoTCP} failure rates for \emph{Probe} resolvers are significantly higher.
As such, \ac{DoTCP} queries targeting \emph{Probe} resolvers fail in 3 out of 4 cases, and, therefore, do not comply with the standard.

With respect to the largest \acp{AS} in terms of probes, we find that failure rates over \ac{DoTCP} for most pairings of \acp{AS} and \emph{Public} resolvers are low, roughly matching the respective failure rates over \ac{DoUDP}.
However, our observation also hints at path-specific issues between the COMCAST and ORANGE \acp{AS} and OpenNIC, where nearly all \ac{DoTCP} \emph{AND} \ac{DoUDP} requests fail.
Looking at \emph{Probe} resolvers, we again observe high failure rates over \ac{DoTCP} across all \acp{AS}, indicating that \emph{Probe} resolvers still lack reliable and vast support for \ac{DoTCP}.

\textbf{Response Times.}
Response times over \ac{DoTCP} are highly varying depending on the continent of the \ac{RA} probes location for \emph{Public} and \emph{Probe} resolvers.
Overall, we find \ac{DoTCP} to be slower than \ac{DoUDP} for nearly all pairings of continent and resolver.
However, when considering the response time differences between both protocols, the relative increase over all continents is less than 37\% for most \emph{Public} and \emph{Probe} resolvers.

Moreover, our evaluation shows that \emph{Public} resolver lack \ac{DoTCP} optimization, not offering support for \emph{EDNS0 TCP keepalive} and \emph{\ac{TFO}}, with the latter only being supported by Google.
However, using the \ac{TFO} cookie in subsequent connections to Google was only successful in rare cases: Due to the connection reset following the refused \ac{TFO} cookie, the usage of \ac{TFO} on Google actually increases the response time in the majority of cases.

\textbf{Outline.}
In~\autoref{sec:related-work} we present related work, followed by our methodology and an overview of our dataset in~\autoref{sec:methodology}. We discuss our failure rate analysis in~\autoref{sub:failure-rates}, followed by the response time analysis in~\autoref{sub:response-times}. Limitations and future work are discussed in~\autoref{sub:limitations} before we conclude the paper with~\autoref{sec:conclusion}.

\section{Related Work}
\label{sec:related-work}

\emph{\acf{EDNS(0)}} are now commonly used in the DNS~\cite{dns.evolution,netalyzr}, \eg to add more information to a DNS message, thereby increasing its size.
However, DNS requests and responses exceeding the path \ac{MTU} cause \emph{IP fragmentation}, which can lead to unreachability due to firewalls blocking fragmented IP packets, or failures due to recipients being unable to reassemble them~\cite{rfc7766,rfc8900,dnssec.fragmentation}.
In addition to the operational challenges this poses, several studies have shown that cache poisoning attacks using \emph{IP fragmentation} can modify \ac{DNS} responses~\cite{dns.fragmentation,domain.validation}.

To avoid \emph{IP fragmentation}, multiple proposals exist to restrict the \ac{DoUDP} payload size through \emph{EDNS(0) Buffer Size} values.
These limits should be in ranges of 1,220--1,472~bytes for IPv4 and 1,220--1,452~bytes for IPv6~\cite{Koolhaas.2020,GeoffHuston.2020,ietf-dnsop-avoid-fragmentation}.
Most notably, the \emph{DNS flag day 2020}~\cite{DNSflagday.2020} proposed a limit of 1,232~bytes for both IPv4 and IPv6, based on the minimum IPv6 \ac{MTU} of 1,280~bytes.

Recent studies have shown that \emph{EDNS(0) Buffer Sizes} of 1,232~bytes or less are already widely used in requests from recursive resolvers to authoritative servers, although 512 and 4,096~bytes remain the predominant sizes~\cite{dns.centralization,truncation}.
While buffer sizes of 512~bytes pose the risk to trigger a \emph{truncation} early, 4,096~bytes exceed the predominant path \ac{MTU} of 1,500~bytes in the Internet~\cite{rfc7766} and, therefore, pose a risk for \emph{IP fragmentation}.
With DNS response sizes becoming larger in general~\cite{nlnet.dnssec,rssac.data.icann}, \emph{truncation} as well as \emph{IP fragmentation} rates will likely increase, and ultimately lead to increased \ac{DoTCP} usage in the future.

Several studies investigate the effects of \emph{truncation}~\cite{truncation,dnssec.fragmentation,netalyzr.implications} and \emph{IP fragmentation}~\cite{truncation,dns.fragmentation,ietf-dnsop-avoid-fragmentation,domain.validation} on \ac{DNS}, although they do not focus on \ac{DoTCP} in detail.
A general recommendation is to prefer \ac{DoTCP} over \ac{DoUDP} in order to avoid \emph{IP fragmentation} in the \ac{DNS}~\cite{ietf-dnsop-avoid-fragmentation,rfc8900,DNSflagday.2020}.
However, a recent study~\cite{dotcp.vulnerable} shows that ICMP messages can be leveraged to trigger \emph{IP fragmentation} on \ac{DoTCP} as well, thereby questioning this recommendation.

Other related work studies the adoption and performance of novel \ac{DNS} protocols like \ac{DoT}~\cite{dot.doh.doudp,dot.doh,dot.doh.2.,TrinhVietDoan.2021} as well as \ac{DoH}~\cite{dot.doh.doudp,dot.doh,dot.doh.2.}, and how they compare to \ac{DoUDP}~\cite{dot.doh.doudp,TrinhVietDoan.2021} in terms of \emph{Failure Rates} and \emph{Response Times}.
However, researchers as well as operators express concerns due to both protocols aiding the trend of Internet centralization~\cite{dns.centralization,dns.centralization.viet}.
Nevertheless, both \ac{DoT} and \ac{DoH} inherently solve the issues of \emph{truncation} and \emph{IP fragmentation} in the \ac{DNS}, yet, their adoption by recursive resolvers is still low~\cite{dot.doh,dot.doh.2.,TrinhVietDoan.2021}.

Considering these evolutions in the DNS, \ac{DoTCP} is needed to prevent \emph{truncation} and \emph{IP fragmentation} until \ac{DoT} or \ac{DoH} are more widely adopted.
While \ac{DoTCP} should already be usable to date as required by the standard~\cite{rfc7766}, the effects of its usage in terms of \emph{Failure Rates} and \emph{Response Times} are not yet extensively studied.

\section{Methodology and Dataset}
\label{sec:methodology}

To study \ac{DoTCP} from the Edge, we perform distributed \ac{DNS} measurements using the \acf{RA} platform~\cite{ripencc:ipj:2015}.

\subsection{Measurement Design}
\label{sub:measurement-design}

\textbf{Measurement Probes.}
We select \ac{RA} hardware probes with the \emph{home} user tag excluding \emph{Anchor} probes: The \emph{home} tag is used to identify \ac{RA} probes operating in residential home networks, of which we find 3,364 probes.
Additionally, we only select probes using hardware version 3 or later due to possible load issues~\cite{Bajpai15,holterbach:imc:2015}.
Of the remaining 2,815 probes, we select 2,500 probes randomly for our measurement study.
From these targeted 2,500 probes, 2,363 probes ultimately execute the measurements (of which 2,361 include location information); the remaining probes were offline or were not considered by the \ac{RA} scheduler during the measurement period.
The final set of probes is distributed across the globe in 83 different countries and 655 distinct \acp{AS}.
Table~\ref{tab:probe-dist-merged} shows the absolute and relative number of the probes per continent and per AS for the top 10 \acp{AS} based on the number of probes.
The relative number is calculated based on 2,361 probes with location information for the continent-based analysis, whereas the AS-based analysis covers 2,363 probes in total.
Note that the locations of the probes are biased toward \ac{EU} and \ac{NA}, as most of the used probes (88.01\%) are deployed in these continents. Similarly, the top 10 source \acp{AS} connect more than a quarter of the used probes (27.04\%) and are also biased towards \ac{EU} and \ac{NA}: While COMCAST, ATT, and UUNET are \ac{NA}-based, the remaining 7 \acp{AS} are \ac{EU}-based.
Hence, our observations are limited by the probes' locations and networks (see~\autoref{sub:limitations}).

\begin{table}[t]
    \centering
    \caption{Distribution of 2,361 RIPE Atlas probes by geographical location (continent) of the probes (top) and by AS for the top 10 ASes based on number of probes (bottom).}
    \begin{tabularx}{\linewidth}{lXr}
        \toprule
        \multicolumn{1}{l}{\textbf{Type}}
        & \multicolumn{1}{X}{\textbf{Location}}
        & \multicolumn{1}{r}{\textbf{Number of Probes}}  \\ 
        \midrule
            & Europe (EU) & 1,636 (69.29\%)                                                 \\
            & \cellcolor[gray]{0.9}North America (NA) & \cellcolor[gray]{0.9}442 (18.72\%) \\
            & Asia (AS)                             & 149 (6.31\%)                         \\
            & \cellcolor[gray]{0.9}Oceania (OC)     & \cellcolor[gray]{0.9}71 (3.01\%)     \\
            & South America (SA)                    & 33 (1.40\%)                          \\
        \multirow{-6}{*}{Continent} 
            & \cellcolor[gray]{0.9}Africa (AF)      & \cellcolor[gray]{0.9}30 (1.27\%)     \\
            \midrule
            & DTAG (AS3320)                         & 127 (5.37\%)                         \\
            & \cellcolor[gray]{0.9}COMCAST (AS7922) & \cellcolor[gray]{0.9}96 (4.06\%)     \\
            & VODANET (AS3209)                      & 93 (3.94\%)                          \\
            & \cellcolor[gray]{0.9}PROXAD (AS12322) & \cellcolor[gray]{0.9}89 (3.77\%)     \\
            & ORANGE (AS3215)                       & 70 (2.96\%)                          \\
            & \cellcolor[gray]{0.9}ATT (AS7018)     & \cellcolor[gray]{0.9}39 (1.65\%)     \\
            & UUNET (AS701)                         & 35 (1.48\%)                          \\
            & \cellcolor[gray]{0.9}TNF (AS33915)    & \cellcolor[gray]{0.9}33 (1.40\%)     \\
            & NTL (AS5089)                          & 29 (1.23\%)                          \\
            & \cellcolor[gray]{0.9}IBSNAZ (AS3269)  & \cellcolor[gray]{0.9} 28 (1.18\%)    \\
        \multirow{-11}{*}{\begin{tabular}[c]{@{}l@{}}Autonomous\\ System\end{tabular}} & others                                      & 1,724 (72.96\%)                      \\
        \bottomrule
    \end{tabularx}
    \label{tab:probe-dist-merged}
\end{table}

\textbf{Public Resolvers.}
For the measurement targets, we select 10 \emph{Public} recursive resolvers based on their usage in related work (see~\autoref{sec:related-work}), querying the IPv4 anycast addresses listed in Table~\ref{tab:resolvers}.

\begin{table}[t]
	\centering
    \caption{Evaluated \textit{Public Recursive Resolvers} and their queried IPv4 Anycast Addresses}
    \label{tab:resolvers}
    \begin{tabularx}{\linewidth}{Xl}
        \toprule
        \textbf{Public Recursive Resolver} & \textbf{IPv4 Anycast Address} \\
        \midrule
        CleanBrowsing & 185.228.168.9 \\
        \rowcolor[gray]{.9}
        Cloudflare DNS & \texttt{1.1.1.1} \\
        Comodo Secure DNS & 8.26.56.26 \\
        \rowcolor[gray]{.9}
        Google Public DNS & \texttt{8.8.8.8} \\
        Neustar UltraDNS & 64.6.64.6 \\
        \rowcolor[gray]{.9}
        OpenDNS & 208.67.222.222 \\
        OpenNIC & 185.121.177.177 \\
        \rowcolor[gray]{.9}
        Quad9 & 9.9.9.9 \\
        UncensoredDNS & 91.239.100.100 \\
        \rowcolor[gray]{.9}
        Yandex.DNS & 77.88.8.8 \\
        \bottomrule
    \end{tabularx}
    \vspace{-2em}
\end{table}


\textbf{Probe Resolvers.}
In addition to the public recursive resolvers, we also query the recursive resolvers configured locally on the probes for comparison.
While every \ac{RA} probe can be configured with multiple resolvers, every DNS measurement which sets the \texttt{use\_probe\_resolver}~option is issued to \emph{every} locally configured resolver.
Hence, one \ac{DNS} measurement request might result in more than one result per probe; the average number of configured resolvers per probe was 2.1.
Based on the source IP address of the DNS responses, we exclude any known unicast or anycast IP address used by the \emph{Public} resolvers listed in Table~\ref{tab:resolvers} for a more accurate distinction and comparison:
The final set is denoted as \emph{Probe} resolvers and represents \ac{ISP} as well as alternative public \ac{DNS} services.

\textbf{DNS Queries.}
We issue \ac{DNS} queries for \texttt{A} records using \ac{DoTCP} as well as \ac{DoUDP}.
For these queries, we include the \emph{EDNS(0)} OPT \ac{RR} with an \emph{EDNS(0) Buffer Size} of 1,232~bytes as proposed by the \emph{DNS flag day 2020}~\cite{DNSflagday.2020}, thereby signaling support to receive DNS responses of up to 1,232~bytes.

To cover a diversity of different domains, which allows us to investigate possible differences between their popularity and region, we construct a set of 200 domains from the Alexa Global and Country Top lists as of March 29, 2021~\cite{alexa.top-sites}.
We sample 150 popularity-focused domains by splitting the Global Top 1M list into 10 evenly-sized bins of 100k each (by rank order) and select the 15 highest-ranked domains of each bin.
For the remaining 50 domains, we determine the countries with the highest numbers of deployed \ac{RA} probes, additionally considering all continents.
For each of those countries (BR, DE, GB, IT, JP, NL, NZ, RU, US, ZA), we choose the 5 highest-ranked domains of the associated country code Top-Level Domain (\eg \texttt{.br}, \texttt{.de}, \texttt{.co.uk}), ultimately resulting in 50 region-focused domains.
Nevertheless, as we do not find any significant deviations for \emph{Failure Rates} or \emph{Response Times}, we do not further distinguish between the domains in our analyses.

To counter limitations of a) the \ac{RA} platform, which enforces the \emph{Recursion Desired} bit on measurements to \emph{Probe} but not to \emph{Public Resolvers}~\cite{ripe.rd.bit}, as well as b) \emph{Public} resolvers utilizing different caching strategies that cannot ensure cached records when using the \ac{RA} platform ~\cite{Randall.2020}, we issue queries to be resolved recursively in order to ensure uncached \ac{DNS} responses, which enables comparable results of \emph{Public} and \emph{Probe} resolvers.
This is achieved by two means:

\begin{enumerate}
    \item \emph{Unique Prefixes}.
    We add \emph{Unique Prefixes} to our 200 domains, which consist of the probe ID and the timestamp of the \ac{DNS} request.
    \item \emph{Recursion Desired}.
    For \emph{Probe Resolver} measurements, the \emph{Recursion Desired} (RD) bit is set by default and enforced on RIPE Atlas for privacy protection~\cite{ripe.rd.bit}.
    However, the bit is \emph{NOT} set by default for \emph{Public} resolver measurements, so we explicitly set the RD bit for all measurements.
\end{enumerate}

While (1) ensures that the queried domain does not exist and is, therefore, not cached, (2) ensures that the resolver will recursively resolve the requested domain.
If \emph{Recursion Desired} would \emph{NOT} be set, a query would \emph{NOT} be recursively resolved but instead be directly responded to by the resolver, even if the queried domain was \emph{NOT} cached or a wildcard matched the queried domain~\cite{rfc8499,Randall.2020}.
Hence, setting \emph{Recursion Desired} on all measurements is required to compare \emph{Public} to \emph{Probe} resolvers.
Moreover, as the overhead of the authoritative resolver lookup is identical on both \ac{DoTCP} and \ac{DoUDP}, the overhead is canceled out for both protocols when analyizing the \emph{differences} in \emph{Response Times} as presented in~\autoref{sub:response-times}, which enables the comparison of \ac{DoTCP} and \ac{DoUDP} measurements.

\textbf{Ethical Considerations.}
The measurement study does not raise any ethical concerns, as we exclusively use RIPE Atlas probes hosted by volunteers, who explicitly agree with publication of the collected data through the RIPE Atlas Service Terms and Conditions~\cite{ra.tos}.
We further query unique domains and set the \emph{RD} bit to ensure recursive name resolution and to avoid \emph{cache snooping}.
Moreover, we aggregate the collected data for the analyses and do not discuss individual probes (or their IP addresses or location coordinates) to preserve the privacy of the voluntary probe hosts.

\textbf{Reproducibility.}
In order to enable the reproduction of our findings~\cite{reproducability}, we make the raw data of our measurements as well as the analysis scripts and supplementary files publicly available on GitHub\footnote{\textbf{\url{https://github.com/kosekmi/2022-ccr-dns-over-tcp-from-the-edge}}}.
Please also refer to the appendix for detailed instructions.

\subsection{Dataset Overview}
\label{sub:dataset-overview}

\begin{table}[t]
    \centering
    \caption{Dataset Overview: \emph{Sample Sizes}, \emph{Failure Rates}, \emph{Failure Reasons}, and \emph{EDNS(0) Buffer Sizes} for \emph{Public} and \emph{Probe} recursive resolvers for \ac{DoTCP} and \ac{DoUDP}.}
    \label{tab:msm-meta-table}
    \resizebox{\linewidth}{!}{\begin{tabular}{lrrrr}
            \toprule
            {} 
            & \multicolumn{1}{c}{\textbf{\begin{tabular}[c]{@{}c@{}}Public\\ DoTCP\end{tabular}}}
            & \multicolumn{1}{c}{\textbf{\begin{tabular}[c]{@{}c@{}}Public\\ DoUDP\end{tabular}}}
            & \multicolumn{1}{c}{\textbf{\begin{tabular}[c]{@{}c@{}}Probe\\ DoTCP\end{tabular}}}
            & \multicolumn{1}{c}{\textbf{\begin{tabular}[c]{@{}c@{}}Probe\\ DoUDP\end{tabular}}}\\ 
            
            \midrule
            \textbf{Samples}                                                             &                      &                      &                      &                      \\ 
            \rowcolor[gray]{.9}
            --- \texttt{Total}                                                           & 4,655,635            & 4,656,086            & 454,151              & 454,417              \\ 
            --- \texttt{Successful}                                                      & 4,282,559            & 4,279,568            & 113,728              & 447,009              \\ 
            \rowcolor[gray]{.9}
            --- \texttt{Failure Rate}                                                    & 8.01\%               & 8.09\%               & 74.96\%              & 1.63\%               \\ 
            
            \midrule 
            \textbf{Failure Reasons}                                                     &                      &                      &                      &                      \\ 
            \rowcolor[gray]{.9}
            --- \texttt{TUCONNECT}                                                       &  4.79\%              &  -                   & 74.72\%              &  -                   \\
            --- \texttt{Timeout}                                                         &  3.22\%              &  8.09\%              &  0.24\%              &  1.63\%              \\
            \rowcolor[gray]{.9}
            --- \texttt{Socket}                                                          &  -                   &  >0.01\%             &  -                   &  -                   \\
            --- \texttt{other}                                                           &  >0.01\%             &  -                   &  -                   &  -                   \\

            \midrule 
            \multicolumn{2}{l}{\textbf{\begin{tabular}[l]{@{}l@{}}EDNS(0) Buffer Sizes\end{tabular} in bytes}}  &                      &                      &                      \\ 
            \rowcolor[gray]{.9}
            --- \texttt{512}                                                             & 26.52\%              & 26.35\%              & 42.11\%              & 28.43\%              \\ 
            --- \texttt{1232}                                                            & 37.51\%              & 44.73\%              & 31.60\%              & 29.93\%              \\ 
            \rowcolor[gray]{.9}
            --- \texttt{4096}                                                            & 35.59\%              & 28.22\%              & 21.61\%              & 27.61\%              \\ 
            --- \texttt{other}                                                           &  0.29\%              &  0.31\%              &  3.34\%              &  8.17\%              \\ 
            \rowcolor[gray]{.9}
            --- \texttt{none}                                                            &  0.09\%              &  0.39\%              &  1.33\%              &  5.87\%              \\ 

            \bottomrule
        \end{tabular}
    }
    \vspace{-2em}
\end{table}


\textbf{Dataset Preparation.}
Overall, we issue a total of 12.1M \ac{DNS} queries (2,500 \emph{probes} $\times$ (10 \emph{Public Resolvers} $+$ on avg. 2.1 \emph{Probe Resolvers}) $\times$ 200 \emph{domains} (with 1 query per domain) $\times$ 2 \emph{Protocols with DoUDP and DoTCP}) as part of our measurement study in April 2021.
As stated in~\autoref{sub:measurement-design}, 2,363 probes execute the measurements and remain in the analysis dataset.
While we explicitly state the IPv4 addresses to be used by our requests to \emph{Public} resolvers, recall that requests to \emph{Probe} resolvers are issued to every locally configured resolver, hence, also over IPv6.
As we focus on IPv4 exclusively in this paper, we leave a comparative study between IPv4 and IPv6 open for future work.
Thus, we exclude all measurements with IPv6 destination addresses (17,556 samples).

In total, we take 4,655,635 \emph{Public} \ac{DoTCP}, 4,656,086 \emph{Public} \ac{DoUDP}, 454,151 \emph{Probe} \ac{DoTCP}, and 454,417 \emph{Probe} \ac{DoUDP} samples into account for our analyses (see Table~\ref{tab:msm-meta-table}).

\textbf{EDNS(0) Buffer Sizes.}
The \emph{EDNS(0) Buffer Size} option allows a \ac{DoUDP} packet to extend its size beyond the default 512~bytes~\cite{rfc1035}, where the signaled buffer size should represent the maximum UDP payload size which the network of the sender can handle~\cite{rfc2671,rfc6891}.
For our queries, we include an \emph{EDNS(0) Buffer Size} (see~\autoref{sub:measurement-design}) in order to check whether the \emph{Public} and \emph{Probe} recursive resolvers support extended buffer sizes through \emph{EDNS(0)}.
If supported, the recursive resolver signals \emph{its} maximum \emph{EDNS(0) Buffer Size} back to the requestor, \ie the maximum UDP payload size which the resolver's network stack should be able to process.
While the resolver knows both the maximum \emph{EDNS(0) Buffer Size} of the requestor as well as its own, the resolver should use the minimum of both signaled values for the actual DNS response so that both endpoints can process the packets accordingly.
Nevertheless, as the signaled \emph{EDNS(0) Buffer Sizes} only represent the maximum buffer sizes that the endpoints should support, the actual size of the response can still exceed the path \ac{MTU}.
Moreover, the \emph{EDNS(0) Buffer Size} is often configured manually~\cite{unbound,bind}, defaulting to sizes which might not be supported by the network in the first place.

Recent work (see~\autoref{sec:related-work}) studies the usage of \emph{EDNS(0) Buffer Sizes} for DNS requests issued from recursive resolvers to authoritative servers~\cite{dns.centralization,truncation}.
In these studies, the authors also observe the rate of \ac{DoTCP} usage on their authoritative servers vantage points:
In the first study~\cite{dns.centralization}, the authors focus on \ac{DNS} cloud providers and find that \ac{DoTCP} is used in up to 15\% of requests issued by Facebook.
In comparison, other evaluated providers (Amazon, Cloudflare, Google, Microsoft) show a usage of only 5\% or below as of 2020.
Similarly, the second study~\cite{truncation} shows a \ac{DoTCP} usage of around 3--5\% of requests as of 2020.
In addition, the authors also evaluate constructed responses of authoritative servers to stub resolvers with a \ac{DoUDP} size of 1,744~bytes, finding that 6.9\% of responses and 3.9\% of probes timed out, and, thus, lead to \ac{DoTCP} fallback.

While we do not control the actual size of the \ac{DNS} responses, we are not able to quantify the actual occurrence of \ac{DoTCP} fallback on responses from recursive to stub resolvers~(see~\autoref{sub:limitations}).
However, our observations complement the aforementioned related studies (between recursive resolvers and authoritative servers) by presenting the signaled \emph{EDNS(0) Buffer Sizes} for \ac{DoUDP} requests issued from recursive to stub resolvers, for which we observe a similar distribution.
Since the buffer sizes stated in Table~\ref{tab:msm-meta-table} are only relevant for \ac{DoUDP}, we did not analyze the observed differences between \ac{DoUDP} and \ac{DoTCP}.
Hence, we detail our observations using \ac{DoUDP} in the following, but include \ac{DoTCP} for completeness.

We measure that 44.73\% of \ac{DoUDP} requests issued to \emph{Public} resolvers and 29.93\% of \ac{DoUDP} requests issued to \emph{Probe} resolvers respond with a buffer size of 1,232~bytes, which complies with the suggested value of the \emph{DNS flag day 2020}~\cite{DNSflagday.2020} and also honors the limits discussed by other proposals~\cite{Koolhaas.2020,GeoffHuston.2020,ietf-dnsop-avoid-fragmentation}.
Notably, \emph{unbound}~\cite{unbound} as well as \emph{BIND9}~\cite{bind} changed their default \emph{EDNS(0) Buffer Sizes} to 1,232~bytes following the \emph{DNS flag day} in 2020. 
However, 26.35\% of \emph{Public} and 28.43\% of \emph{Probe} resolvers respond with a buffer size of 512~bytes, and 28.22\% of \emph{Public} and 27.61\% of \emph{Probe} resolvers respond with 4,096~bytes.

Most \emph{Public} resolvers use a single \emph{EDNS(0) Buffer Size} predominantly (>95\%).
Cloudflare, UncensoredDNS, and Yandex primarily use 1,232~bytes, while Comodo and OpenDNS use 4,096~bytes.
On the other hand, CleanBrowsing and Google mainly use 512~bytes, whereas OpenNIC (75.4\% with 4,096~bytes; 23.8\% with 1,232~bytes), Quad9 (48.5\% with 1,232~bytes; 49.8\% with 512~bytes), and Neustar (50.2\% with 1,232~bytes; 49.0\% with 4096~bytes) show a mixed usage of buffer sizes instead.

Notably, our observations on Google responding with an \emph{EDNS(0) Buffer Size} of 512~bytes in 98.0\% of cases does differ to the observation made by~\cite{dns.centralization}.
The authors find that 24\% of requests from Google to authoritative servers use \emph{EDNS(0) Buffer Sizes} of up to 1,232~bytes, with the remaining 76\% primarily using 4,096~bytes instead.
Mapping our results to these observations shows that Google uses different \emph{EDNS(0) Buffer Sizes} on stub-facing resolvers in comparison to authoritative-facing resolvers.

Other \emph{EDNS(0) Buffer Sizes} are seen in 0.31\% of cases for \emph{Public} resolvers, and 8.17\% of cases for \emph{Probe} resolvers.
\emph{Public} resolvers show no buffer size greater than 4,096~bytes; in contrast, \emph{Probe} resolvers exceed this value in 0.95\% of cases with buffer sizes of 8,192~bytes (0.34\%), 65,494~bytes (0.58\%), and 65,535~bytes (0.03\%).

Our observations show, that \ac{DNS} responses from recursive to stub resolvers use \emph{EDNS(0) Buffer Sizes} of 512 and 4,096~bytes in more than 55\% of cases, thereby falling considerably short of or exceeding the recommended limits.
While these results allow us to put our observations into perspective, a comprehensive study on \emph{truncation} and \emph{IP fragmentation} for requests issued from stub to recursive resolvers is left for future work (see~\autoref{sub:limitations}).

\vspace{-1em}
\section{Failure Rates}
\label{sub:failure-rates}

In order to assess the reliability of \ac{DoTCP}, we study the number of failures which the probes observe during their measurements.
We define a measurement as failed if the probe did NOT receive a response from the queried resolver, and state the failure reasons according to the data provided by \acf{RA}: either due to issues with the TCP connection (\emph{TUCONNECT}), with receiving a DNS response within 5 seconds (\emph{Timeout}), or with sending the DNS request (\emph{Socket}).
We then determine the \emph{Failure Rate} and \emph{Failure Reasons} as the relative number of failures based on all measurements.
Table~\ref{tab:msm-meta-table} lists the overall \emph{Failure Rates} and \emph{Failure Reasons} by \emph{Public} and \emph{Probe} resolver measurements for both \ac{DoTCP} and \ac{DoUDP}.

\emph{Public} resolvers exhibit similar \emph{Failure Rates} for both \ac{DoTCP} (8.01\%) and \ac{DoUDP} (8.09\%), showing that the reliability of \ac{DoTCP} is comparable to that of \ac{DoUDP}.
In terms of \emph{Failure Reasons}, almost all failures of \ac{DoUDP} measurements to \emph{Public} resolvers are attributed to \emph{Timeout}, whereas \ac{DoTCP} measurements to \emph{Public}  resolvers show \emph{TUCONNECT} as the primary failure reason with 4.79\%.

Note that \ac{RA} does not provide more detail on \emph{TUCONNECT} errors; previous work~\cite{TrinhVietDoan.2021} on \ac{DoT} measurements using \ac{RA} suggested that these failures are related to TLS negotiation errors.
However, since we observe this behavior using TCP as well, it is more likely that \emph{TUCONNECT} hints at issues with the (underlying) TCP connection instead, \ie the \ac{RA} probe is not able to establish a TCP connection with the recursive resolver (which in return causes a potential TLS negotiation to also fail).

More specifically, we attribute \emph{TUCONNECT} to instances where the probe is informed about the unreachability of the contacted IP:Port combination by receiving a TCP \texttt{RST} to the probes TCP \texttt{SYN} packet, hence stating that the recursive resolver does not support \ac{DoTCP}.
On the other hand, a \emph{Timeout} is recorded if loss occurs, or if no TCP \texttt{RST} is received, either due to not being elicited in the first place or due to being lost in transit.
To substantiate this hypothesis, we issue \ac{RA} \ac{DoTCP} requests targeting a controlled recursive resolver: Thus, we are able to verify that elicited TCP \texttt{RST} packets result in \emph{TUCONNECT}, whereas \ac{RA} reports \emph{Timeout} if no packet was sent in response to the TCP \texttt{SYN}.
Please note that both failures might also occur if middleboxes drop the request, either silently (resulting in \emph{Timeout}), or by eliciting a TCP \emph{RST} packet themselves (resulting in \emph{TUCONNECT}).
Since we measure \ac{DoTCP} from the edge, we cannot analyze possible path influences in more detail (see~\autoref{sub:limitations}).

Evaluating \emph{Probe} resolvers, a significantly different behavior is shown in comparison to \emph{Public}:
For \ac{DoUDP}, \emph{Probe} resolvers have a fairly low failure rate with 1.63\%.
However, measurements attempting \ac{DoTCP} fail in 74.96\% of the cases, with \emph{TUCONNECT} accounting for 74.72\% (remaining 0.24\% \emph{Timeout}), which indicates that vast support for \ac{DoTCP} among \emph{Probe} resolvers is lacking.

In particular, RFC~7766~\cite{rfc7766} states that implementations of authoritative servers, recursive resolvers, and stub resolvers MUST support \ac{DoTCP}.
In addition to the dominance of the \emph{TUCONNECT} failure reason, the high failure rate of \emph{Probe} resolvers for \ac{DoTCP} in contrast to \emph{Public} resolvers also indicates that these failures occur due to missing \ac{DoTCP} support on the side of \emph{Probe} resolvers.
We suspect that this is due to most \emph{Probes} using \ac{CPE} devices (e.g., home routers) as their resolvers, which typically forward DNS queries to an upstream DNS service operated by the \ac{ISP} \cite{schomp.2013}:
Out of the 74.96\% \ac{DoTCP} measurements that failed for the \emph{Probe} resolvers (see Table~\ref{tab:msm-meta-table}), 99.47\% were issued to resolvers with private IPv4 addresses (\ie 74.56\% of all \ac{DoTCP} measurements to \emph{Probe} resolvers).
Thus, the observation indicates that almost all of the measurements to \emph{Probe} resolvers are forwarded from \ac{CPE} devices to ISP resolvers which do not implement \ac{DoTCP}, and therefore do not comply with the standard defined by RFC~7766~\cite{rfc7766}.

\takeaway{While we find failure rates over \ac{DoTCP} to be comparable with \ac{DoUDP} for \emph{Public} resolvers, \ac{DoTCP} failure rates for \emph{Probe} resolvers are significantly higher.
As such, \emph{Probe} resolvers cannot successfully return large DNS responses that require a fallback to \ac{DoTCP} in 3 out of 4 cases, and, thus, do not comply with the standard.}

\begin{figure*}[t]
	\centering
	\includegraphics[width=\linewidth]{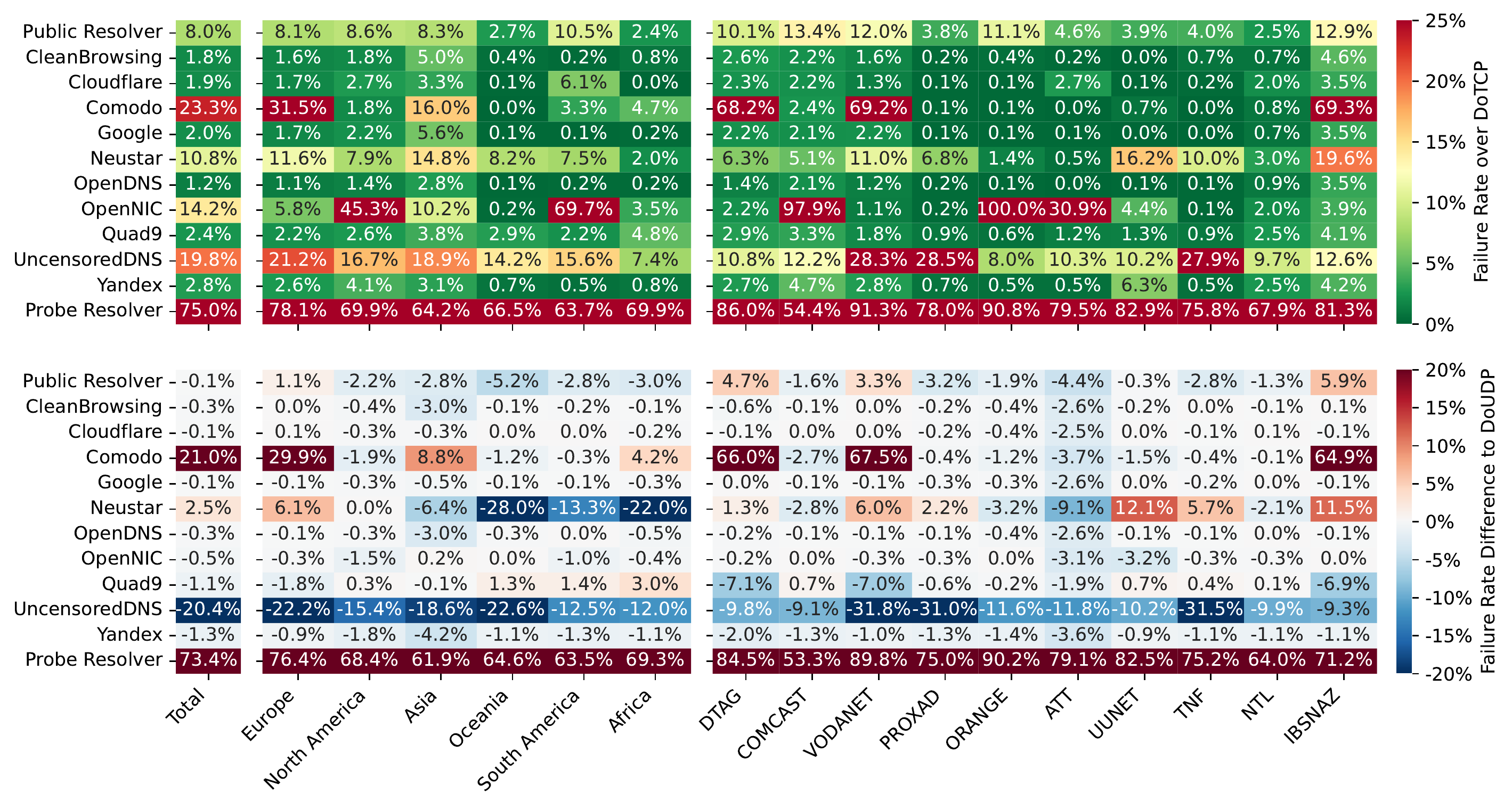}
	\caption{Failure rate by Resolver over \ac{DoTCP} (top), along with respective failure rate difference in percentage points between \ac{DoTCP} and \ac{DoUDP} (bottom); across all samples in total (left), by Continent (middle), and by Top 10 ASes (right), each in descending order by number of probes. Positive values (colored in red) indicate higher failure rates for \ac{DoTCP}.
    } \label{fig:heatmap-fr}
	\vspace{-1em}
\end{figure*}

\paragraph{\textbf{By Continent}}
In order to investigate regional differences for \ac{DoTCP}, we group the failures for each resolver and continent (based on geographic coordinates pulled from the \ac{RA} probe API), and calculate the respective \emph{Failure Rates}, as shown in Fig.~\ref{fig:heatmap-fr} (top). While the top row aggregates all \emph{Public} resolvers, each of the 10 \emph{Public} resolvers as well as the \emph{Probe} resolvers are detailed in the remaining rows. Note that this layout is the same for all remaining (sub)plots in the paper.

We find that \ac{DoTCP} \emph{Failure Rates} vary between different resolvers, as the total \emph{Failure Rates} across all continents (top left) are within the range of 1.2--2.8\% for CleanBrowsing, Cloudflare, Google, OpenDNS, Quad9, and Yandex.
On the other hand, Comodo, Neustar, OpenNIC, and UncensoredDNS show considerably higher \emph{Failure Rates} with 10.8--23.3\%.
Notably, Comodo and UncensoredDNS also show \emph{TUCONNECT} as the primary \emph{Failure Reason} with 95\% and 92\% of failures, respectively.
In contrast, we observe mixed \emph{Failure Reasons} with comparable occurrences of \emph{TUCONNECT} and \emph{Timeout} for the remaining \emph{Public} resolvers.
This indicates that \ac{DoTCP} is not offered universally on all \acp{PoP} of the \emph{Public} resolvers, as our observations show no clear preference for one specific failure reason.

Further, \emph{Failure Rates} for a specific resolver also differ between continents (Fig.~\ref{fig:heatmap-fr} top middle): E.g., we observe higher \emph{Failure Rates} for most resolvers in \ac{SA} and \acf{AS}.
\emph{Probe} resolvers have \emph{Failure Rates} of 63.7--78.1\% across all continents, with resolvers in \ac{SA} showing the lowest \emph{Failure Rates}.
We also find outliers in \ac{EU} and \ac{NA}, where Comodo (EU: 31.5\%) and OpenNIC (NA: 45.3\%) have significantly higher \emph{Failure Rates} for \ac{DoTCP}; OpenNIC further exhibits a higher failure rate in \ac{SA} with 69.7\%.

Comparing \ac{DoTCP} and \ac{DoUDP}, Fig.~\ref{fig:heatmap-fr} (bottom) presents the absolute \emph{Failure Rate} differences (in percentage points) between the \ac{DoTCP} and \ac{DoUDP} measurements for the resolvers and continents (bottom middle); \ie positive values (colored in red) indicate higher \emph{Failure Rates} for \ac{DoTCP}, whereas negative values (colored in blue) represent higher \emph{Failure Rates} over \ac{DoUDP} instead.
Overall, the differences between \ac{DoUDP} and \ac{DoTCP} are marginal for the resolvers showing low \ac{DoTCP} failure rates, with the differences being around zero percentage points.
However, probes across all continents (bottom left) experience lower failure rates to UncensoredDNS for \ac{DoTCP} instead of \ac{DoUDP}, as the \ac{DoUDP} \emph{Failure Rates} are higher by 12.0--22.6 percentage points in comparison.
Similarly, measurements to Neustar failed more frequently over \ac{DoUDP} (by 6.4--28.0 percentage points) for all continents except for \ac{EU} and \ac{NA}.
On the other hand, Comodo exhibits much higher \emph{Failure Rates} over \ac{DoTCP} than over \ac{DoUDP} for \ac{AS} (8.8 percentage points) and \ac{EU} (29.9 percentage points), which results in an overall difference of 21.0 percentage points in favor of \ac{DoUDP} across all continents for Comodo.
For \emph{Probe} resolvers, the \ac{DoTCP} \emph{Failure Rates} are also significantly higher: The \emph{Failure Rate} differences range from 61.9 to up to 76.4 percentage points; considering the absolute results described above (see Fig.~\ref{fig:heatmap-fr} top middle), which barely differ from the percentages shown in the subtraction plot (see Fig.~\ref{fig:heatmap-fr} bottom middle), we observe that \ac{DoUDP} is still significantly more reliable in comparison with \ac{DoTCP} across all continents for \emph{Probe} resolvers.

\takeaway{Overall, we find that across nearly all continent and \emph{Public} resolver pairings, \ac{DoTCP} exhibits a roughly similar failure rate in comparison with \ac{DoUDP}. However, not all resolver \acp{PoP} of a \emph{Public} resolver support \ac{DoTCP} universally, resulting in different failure reasons. As for \emph{Probe} resolvers, we observe that failure rates over \ac{DoTCP} are much higher on each continent, ranging from roughly 63\% to 78\%, whereas \ac{DoUDP} is much more reliable for \emph{Probe} resolvers.}

\paragraph{\textbf{By Autonomous System}} 
We further study the failure rates for the largest 10 \acp{AS} (based on number of \ac{RA} probes), i.e., we group the samples by resolver and AS before calculating the failure rates.
Fig.~\ref{fig:heatmap-fr} shows the failure rates over DoTCP by AS (top right) for the subset of 639 probes hosted in the top 10 \acp{AS}, along with the failure rate difference in percentage points to DoUDP (bottom right).
Note that due to the deployment of \ac{RA} probes (see~\autoref{sub:measurement-design}), the top 10 ASes are inherently centered around \ac{EU} and \ac{NA}.

Overall, we observe that failure rates are around <1--3\% for most AS-\emph{Public} resolver pairings.
Similarly, the differences to failure rates over DoUDP are <1\% for most, indicating that both \ac{DoTCP} and \ac{DoUDP} resolvers work fairly reliably with most \emph{Public} resolvers.

The \acp{AS} themselves show comparable failure rates for most \emph{Public} resolvers, while Neustar and UncensoredDNS exhibit increased failure rates ranging from 5.1\% up to 28.5\% for almost all \acp{AS}.
In contrast, we notice some pairings that show significantly higher failure rates:
For instance, we find outliers in failure rates of roughly 68--69\% for DoTCP requests from DTAG, VODANET, and IBSNAZ to Comodo.
As 95\% of all failed \ac{DoTCP} measurements to Comodo are \emph{TUCONNECT} errors, \ac{RA} probes from those \acp{AS} are unable to reliably establish TCP connections with Comodo's recursive resolvers.
This is also reflected in the bottom plot, where the increases of 64.9--67.5 percentage points for the same \acp{AS} and Comodo show that failures are much less common over \ac{DoUDP}.

Moreover, probes hosted in the \acp{AS} of COMCAST and ORANGE experience even higher failure rates with 97.9\% and 100.0\%, respectively, towards OpenNIC; probes in the ATT AS also show moderately high failure rates of 30.9\%.
This indicates that nearly all DoTCP requests from the former two \acp{AS} encounter issues which lead to no valid DoTCP responses from OpenNIC.
We find that 96\% of all failed DoTCP measurements to OpenNIC result in \emph{Timeout} errors which surpass the 5 second threshold.
Considering that other \emph{Public} resolvers and \acp{AS} do \emph{NOT} show similarly high failure rates for OpenNIC, this observation suggests issues specific to the paths between OpenNIC and the \acp{AS} of COMCAST as well as ORANGE, \eg blackholing.
This is supported by the fact that the differences in failure rates for OpenNIC (Fig.~\ref{fig:heatmap-fr} bottom) are 0.0\% for both COMCAST and ORANGE, stating that the failure rates are equal using \ac{DoTCP} and \ac{DoUDP}.

For \emph{Probe} resolvers, we observe failure rates ranging from 54.4\% for COMCAST up to 91.3\% for VODANET across all \acp{AS}.
In contrast to \ac{DoUDP}, failure rates are much higher, as the differences in percentage points shown in the bottom plot are about equally as high, ranging from 53.3 to 90.2 percentage points.
Given that the \emph{Probe} resolver failure rates are even higher than seen in the continent-level analysis, this observation supports our above hypothesis which states that most of the measurements to \emph{Probe} resolvers are forwarded from \ac{CPE} devices to ISP resolvers with lacking \ac{DoTCP} support.

In some cases, we find DoTCP to be more reliable than DoUDP:
While UncensoredDNS exhibits moderate failure rates (between 8.0\% and 28.5\%) across all \acp{AS} as outlined above, we notice that the failure rate difference to \ac{DoUDP} ranges from $-$9.1 to $-$31.8 percentage points.
As such, \ac{DoTCP} is more reliable than \ac{DoUDP} for the top 10 \acp{AS} when using UncensoredDNS, with most of the DoTCP failures being related to \emph{TUCONNECT} errors (92\%).
The same pattern applies to Yandex, which shows failure rates from 0.5\% to 6.3\% for the different \acp{AS} over DoTCP, whereas the differences to \ac{DoUDP} are between $-$0.9 and $-$3.6 percentage points, meaning that \ac{DoTCP} is more reliable than \ac{DoUDP} for Yandex.
Similarly, Quad9 shows failure rate differences of roughly $-$7 percentage points for the DTAG, VODANET, and IBSNAZ \acp{AS} in particular, which means moderately high failure rates of 8--11\% over \ac{DoUDP} but much more reliable \ac{DoTCP} behavior with 1.8--4.1\% failure rates for these \acp{AS}.
In contrast, Neustar samples show varying failure rates over \ac{DoTCP} ranging from 0.5\% to 19.6\%, as well as failure rate differences to \ac{DoUDP} between $-$9.1 and $+$12.1 percentage points.

\takeaway{In our AS-based analysis, we find that failure rates over \ac{DoTCP} for most pairings of \acp{AS} and \emph{Public} resolvers are low \mbox{(<1--3\%)}, which roughly matches the respective failure rates over \ac{DoUDP} (difference around 0 percentage points). However, we also observe cases in which failure rates over \ac{DoTCP} are much higher: For some \acp{AS}, \ac{DoUDP} requests are mostly successfully responded to by Comodo; yet, the \ac{DoTCP} requests failed in more than two out of three measurements from these \acp{AS} (68.2--69.3\% failure rates).
Moreover, our observation also indicate paths specific issues between OpenNIC and COMCAST as well as ORANGE, showing failure rates of 97.9\% and 100.0\% for \ac{DoTCP} \emph{AND} \ac{DoUDP}.
Regarding \emph{Probe} resolvers, we again observe high failure rates over \ac{DoTCP} (54.4--91.3\%) across all top 10 \acp{AS}, indicating that \emph{Probe} resolvers still lack reliable and vast support for \ac{DoTCP}.}

\section{Response Times}
\label{sub:response-times}

\begin{figure*}[t]
	\centering
	\includegraphics[width=\linewidth]{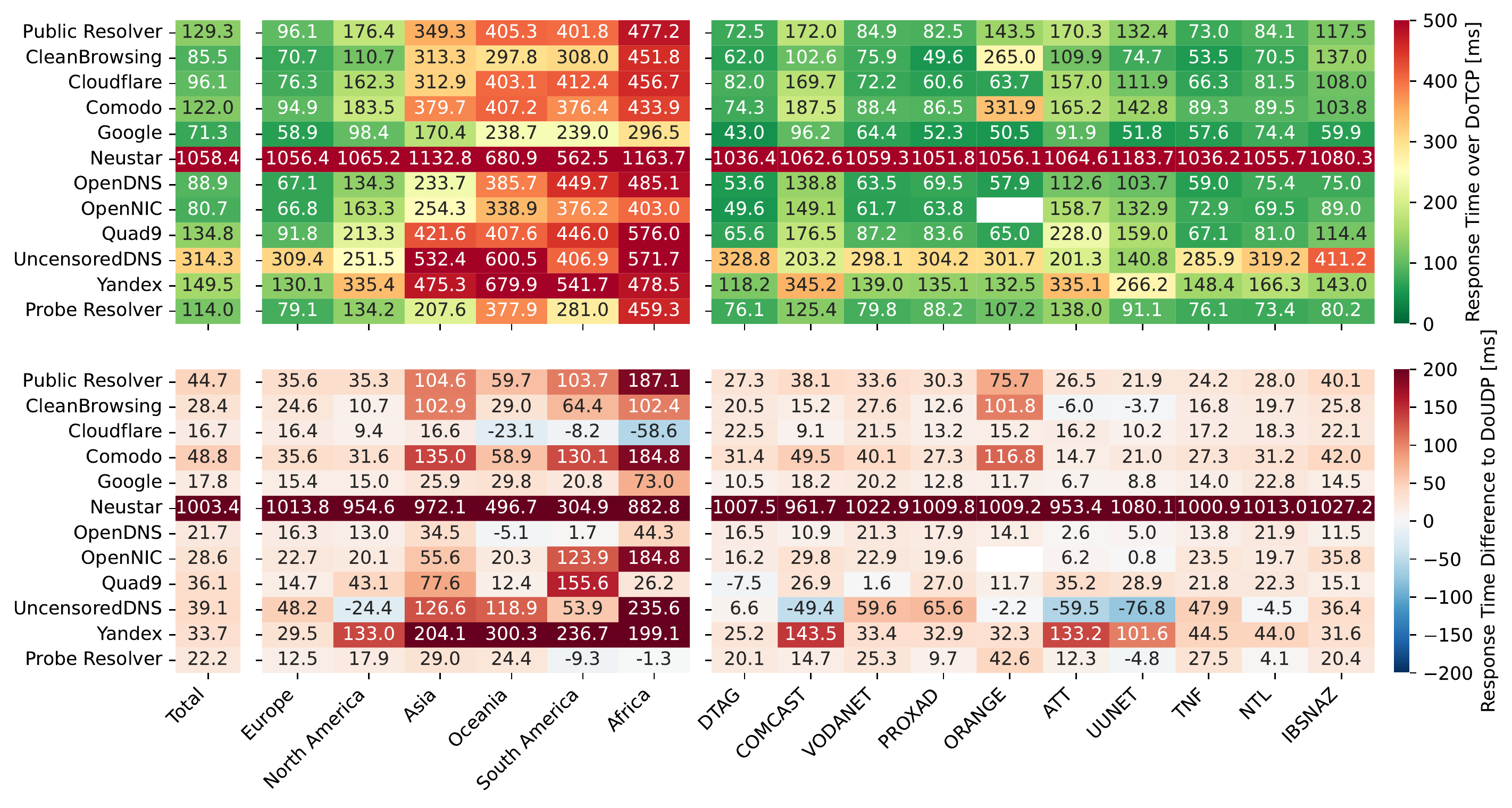}
	\caption{Median response time by Resolver based on medians for probe and resolver over \ac{DoTCP} (top), along with respective response time difference between \ac{DoTCP} and \ac{DoUDP} (bottom); across all samples in total (left), by Continent (middle), and by Top 10 ASes (right), each in descending order by number of probes. Positive values (colored in red) indicate higher median response times for \ac{DoTCP}.
	} \label{fig:heatmap-rt}
	\vspace{-1em}
\end{figure*}

To enable a direct comparison of \ac{DoTCP} and \ac{DoUDP}, we only include probe:resolver pairs with both successful \ac{DoTCP} \emph{AND} \ac{DoUDP} measurements (see Table~\ref{tab:msm-meta-table}) in our \emph{Response Times} analysis.
Moreover, please recall that domain names are explicitly resolved recursively, which ensures uncached \ac{DNS} responses and enables comparable results for \emph{Public} and \emph{Probe} resolvers~(see~\autoref{sec:methodology}).

The \emph{Response Time} is defined as the time between the moment the first packet of the measurement is sent by the \ac{RA} probe until the moment it receives a valid DNS response.
While the first packet for \ac{DoUDP} is the actual DNS query, the TCP 3-way handshake \texttt{SYN} is the first packet for \ac{DoTCP}.
Since we ensure uncached responses, the \emph{Response Time} also includes the time required for the lookup of the requested domain on the authoritative resolver, and therefore comprises in detail of:
\vspace{-0.3em}
\begin{enumerate}
    \item In case of \ac{DoTCP}:
    \\Connection Establishment: Probe $\leftrightarrow$ Recursive
    \item Request: Probe $\rightarrow$ Recursive $\rightarrow$ Authoritative
    \item Response: Authoritative $\rightarrow$ Recursive $\rightarrow$ Probe
\end{enumerate}
Hence, \ac{DoTCP} requires an additional \ac{RTT} for the connection establishment between Probe and recursive resolver (step (1)).
Thus, we expect \ac{DoTCP} to result in higher \emph{Response Times} compared to \ac{DoUDP}. 
Therefore, the \emph{Response Times} over \ac{DoTCP} presented in Fig.~\ref{fig:heatmap-rt} (top) resemble the time required for all steps (1)--(3).
Moreover, the \emph{Response Time} \emph{differences} between \ac{DoTCP} and \ac{DoUDP} shown in Fig.~\ref{fig:heatmap-rt} (bottom) represent the overhead caused by the TCP connection establishment (step (1)), regardless of whether a cached or uncached record is looked up, as the calculated \emph{differences} essentially nullify steps (2)--(3) altogether.

\paragraph{\textbf{By Continent}}
We evaluate the observed \ac{DNS} response times by calculating the median response times per resolver and continent based on the median response times of each probe:resolver pair as shown in Fig.~\ref{fig:heatmap-rt} (top middle).

We observe that \ac{DoTCP} response times vary between resolvers and continents, where Neustar performs considerably worse on each individual continent ranging from 562.5ms in \ac{SA} to 1,163.7ms in \ac{AF}.
In contrast, Google does offer the fastest response times over all continents with 71.3ms, as well as on each individual continent.
Moreover, we find regional differences over all resolvers, where \ac{EU} (58.9--309.4ms) and \ac{NA} (98.4--335.4ms) show considerably faster response times over all resolvers except Neustar. 
Evaluating \ac{AF}, we observe that the continent shows the slowest response times for 8 of the 10 \emph{Public} resolvers ranging from 296.5ms (Google) to 1,163.7ms (Neustar), hinting at fewer \acp{PoP} in \ac{AF}.
To check this hypothesis, we lookup information on the DNS infrastructures published by the operators of the \emph{Public} resolvers~\cite{pops-cleanbrowsing,pops-cloudflare,pops-comodo,pops-google,pops-neustar,pops-opendns,pops-opennic,pops-quad9,pops-uncensoreddns,pops-yandex}; for most resolvers, we find that the number of \acp{PoP} in \ac{AF} is indeed lower than in other continents.

To compare \ac{DoTCP} to \ac{DoUDP} response times, Fig.~\ref{fig:heatmap-rt} (bottom middle) shows the response times \emph{difference} between \ac{DoTCP} and \ac{DoUDP} for all resolvers and continents.
Positive values (colored in red) indicate higher response times for \ac{DoTCP}, and negative values (colored in blue) represent higher response times over \ac{DoUDP}.

In total, the response times increase moderately when using \ac{DoTCP} instead of \ac{DoUDP}, ranging from 16.7ms (\ie an increase by 21.0\%, Cloudflare) to 48.8ms (66.6\%, Comodo) for all \emph{Public} resolvers except Neustar, where the response time is increased by 1,003.4ms (1,824.6\%).
Overall, the relative increase is less than 37\% for 6 out of 10 \emph{Public} resolvers.
Cloudflare does show minor increases with 16.6ms for \ac{AS}, 16.4ms for \ac{EU}, and 9.4ms for \ac{NA}, but does manage to achieve lower response times over \ac{DoTCP} in comparison with \ac{DoUDP} in \ac{AF} ($-$58.6ms), \ac{OC} ($-$23.1ms), and \ac{SA} ($-$8.2ms) as well.
Notably, \emph{Probe} resolvers achieve lower response times for \ac{DoTCP} in \ac{SA} ($-$9.3ms) and \ac{AF} ($-$1.3ms), and also show only minor increases of 12.5--29.0ms for the remaining continents.
This results in a total increase of 22.2ms for all \emph{Probe} resolvers, which is a relative increase of 24.1\% when switching from \ac{DoUDP} to \ac{DoTCP}.
As \emph{Probe} resolvers primarily consist of \ac{ISP} resolvers (see~\autoref{sub:measurement-design} and~\autoref{sub:failure-rates}), they are located closer to the \emph{home} probes than \emph{Public} resolvers which are typically hosted in data centers farther away~\cite{dns.centralization.viet}.
Therefore, we attribute the lower observed response times to these shorter paths, which result in lower latencies due to faster handshakes.

\takeaway{Response times over \ac{DoTCP} are highly varying depending on the continent of the probe for \emph{Public} and \emph{Probe} resolvers, ranging from 58.9ms to more than one second. Response times are especially high in \ac{AF} (296.5--1,163.7ms), which we attribute to the lower number of resolver \acp{PoP} in \ac{AF}. On the other hand, we observe the lowest response times for \ac{EU} and \ac{NA}, which are both continents with the most \acp{PoP} w.r.t. both resolver endpoints and \ac{RA} probes. Nevertheless, we find \ac{DoTCP} to be slower than \ac{DoUDP} for nearly all pairs of continent and resolver, with largely varying response time differences (from $-$58.6ms to 1,013.8ms).
However, the relative increase over all continents is less than 25\% for all \emph{Probe} resolvers, and less than 37\% for 6 out of 10 \emph{Public} resolvers.}

\paragraph{\textbf{By Autonomous System.}}
To investigate response times for the top 10 \acp{AS}, we calculate the median response times per resolver and AS based on the median response times of the probe:resolver pairs, i.e., analogous to the continent-based analysis.
Fig.~\ref{fig:heatmap-rt} presents the absolute medians for each resolver and AS (top right), as well as the difference in response times to DoUDP (bottom right).
Note that due to the 100.0\% failure rate observed for ORANGE using OpenNIC (see~\autoref{sub:failure-rates}), the respective value could not be determined, which is denoted by the empty cell in the plot.

Overall, the response times roughly match the continent-based response time for \ac{EU} and \ac{NA} as discussed in the previous analysis, however, recall that the top 10 \acp{AS} are centered around \ac{EU} and \ac{NA}.
Other patterns, such as Neustar and UncensoredDNS showing higher response times overall, also apply to the AS-based analysis.
Similarly, the determined response times are comparable across most resolvers for an individual AS, although we also see outliers with higher response times:

For instance, the \ac{NA}-based \acp{AS}, namely COMCAST, ATT, and UUNET exhibit higher response times of mostly above 100ms, especially for Yandex (266.2--345.2ms).
This is likely because \acp{PoP} of Yandex' DNS service are located in Russia, Commonwealth of Independent States (CIS) countries, and Western Europe~\cite{pops-yandex}: Therefore, probes of \ac{NA}-based \acp{AS} are located much farther away than probes in \ac{EU}, resulting in significantly higher response times.

In contrast, the remaining \acp{AS}, all of which are \ac{EU}-based, measure response times of approximately 50--90ms for most pairings.
This is likely due to larger geographical distances that packets have to travel in \ac{NA}, compared to the more compact landmass in \ac{EU}, which is ultimately reflected in the \acp{AS} as well. 
An exception to this is IBSNAZ (based in Italy), which shows higher response times of around 100ms for most resolvers, although this is still lower than the response times seen by the \ac{NA}-based \acp{AS}.
On the other hand, samples from ORANGE also show higher response times when querying CleanBrowsing (265.0ms) and Comodo (331.9ms).

Contrasting the response times of \emph{Probe} and \emph{Public} resolvers, the AS-based analysis reveals that \emph{Probe} resolvers are mostly on par with \emph{Public} resolvers, whereas the failure rates are significantly worse (see~\autoref{sub:failure-rates}).

Regarding the response time differences between DoTCP and DoUDP, we observe that for most pairings \ac{DoTCP} is slower by up to 50ms, with many values accumulating around 10--20ms.
However, we find that UncensoredDNS achieves response time differences between $-$49.4ms and $-$76.8ms for the US-based \acp{AS}, indicating that responses from UncensoredDNS over DoTCP are faster than over DoUDP for these \acp{AS}.
Nevertheless, the exceptional cases of higher response times over DoTCP for specific AS-resolver pairings (discussed above) are all reflected through higher deltas in the difference plot as well, overall showing that DoTCP is slower than DoUDP for probes in the top 10 \acp{AS}.

\takeaway{The top 10 \acp{AS} provide a more in-depth perspective of the response time analysis for probes in \ac{EU} and \ac{NA}: We notice that \ac{NA}-based \acp{AS} are slow when using Yandex due to geographical distance, as Yandex primarily operates around Russia. On the flip side, \ac{EU}-based \acp{AS} measure the lowest response times, which we attribute to the high density of both resolver \acp{PoP} and probe deployments. Altogether, we find \emph{Probe} resolvers exhibit roughly comparable response times to \emph{Public} resolvers. Nevertheless, across each AS and nearly all resolvers, \ac{DoTCP} is slower than \ac{DoUDP}.}

\paragraph{\textbf{Optimizations.}} 

To put our results into perspective, we take a closer look at two key features aiming to improve \ac{DoTCP} response time, which are both recommended by recent standardization efforts focussing on operational requirements for \ac{DoTCP}~\cite{ietf-dnsop-dns-tcp-requirements}: \emph{\ac{TFO}} and \emph{EDNS0 TCP keepalive}.
While \emph{\ac{TFO}}~\cite{rfc7413} reduces the handshake time by one \ac{RTT} for connections following an initial exchange of a \ac{TFO} cookie, \emph{EDNS0 TCP keepalive}~\cite{rfc7828} allows \ac{DNS} resolvers to keep a TCP connection alive.
Both mechanisms can be leveraged to reduce the \ac{DoTCP} response time by one \ac{RTT}, hence, bringing it on par with \ac{DoUDP}.
Although \ac{RA} does not provide information on the usage of \emph{\ac{TFO}} or \emph{EDNS0 TCP keepalive} within their documentation (or the measurement results itself), we setup a recursive resolver to explicitly check the probe's support for both features.
For this, we randomly selected 50 probes and issued one \ac{DoTCP} query per probe.
We find that \emph{none} of the requests include either the \emph{\ac{TFO}} or the \emph{EDNS0 TCP keepalive} option.
As \ac{RA} probes issue \ac{DNS} measurements identically with the same options and parameters, we conclude that none of our requests used the features.

To evaluate the general support of these features on the \emph{Public} resolvers, we manually check each of the \emph{Public} resolvers by explicitly requesting a \ac{TFO} cookie and setting the \texttt{edns-tcp-keepalive} \emph{EDNS(0)} option in our queries from a single vantage point.
However, none of the tested resolvers returned the \texttt{edns-tcp-keepalive} \emph{EDNS(0)} option required for \emph{EDNS0 TCP keepalive}, nor a \ac{TFO} cookie required for \emph{\ac{TFO}}.
An exception to this is Google, which responds with a \ac{TFO} cookie.
Using the cookie in subsequent connections, however, was only successful in rare cases:
The resolver terminated the connection upon receiving the \ac{TFO} cookie for most measurements, falling back to a traditional TCP handshake instead.
This behavior is also observed by~\cite{dot.doh} and indicates \ac{DNS} load balancing, which does not factor previous connections into account for server selection.
While~\cite{ietf-dnsop-dns-tcp-requirements} recommends the usage of \ac{TFO} by leveraging the same \ac{TFO} key in load balancing scenarios as well, the behavior as observed on Google actually increases the response time by one \ac{RTT} due to the connection reset following the refused \ac{TFO} cookie.

\takeaway{None of the tested public resolvers support \emph{EDNS0 TCP keepalive}, and \emph{\ac{TFO}} is only supported by Google. However, using the \ac{TFO} cookie in subsequent connections to Google was only successful in rare cases: due to the connection reset following the refused \ac{TFO} cookie, the usage of \ac{TFO} on Google actually increases the response time in the majority of cases.}


\section{Limitations \& Future Work}
\label{sub:limitations}

We acknowledge that observations from \ac{RA} probes are not fully representative and unconditionally generalizable for the whole Internet.
Given probes are mainly deployed in \ac{EU} and \ac{NA}, the probes selected in our study are also heavily centered around these regions (see~\autoref{sub:measurement-design}).
However, we still aggregate \emph{all} samples across a continent (and AS) for the \emph{Failure Rate} and \emph{Response Time} analyses: Reducing the number of samples from \ac{EU} and \ac{NA} to be comparable to other continents would have overall resulted in a much lower number of data points and, therefore, in a reduced representativeness of the measurement study.

Since we measure \ac{DoTCP} from the Edge, we cannot control or analyze possible path influences on the signaled \emph{EDNS(0) Buffer Sizes} (see~\autoref{sub:dataset-overview}).
For instance, middleboxes might change the \emph{EDNS(0) Buffer Size} based on a static configurations or a discovered path \ac{MTU}.
RFC 6891~\cite{rfc6891} explicitly prohibits this for simple DNS forwarders.
However, the RFC makes an exception for middleboxes with additional functionality, which are allowed to process and act on the \emph{EDNS(0) Buffer Size}; \eg \emph{CoreDNS}~\cite{coredns.bufsize} leverages this to override the buffer size in order to prevent \emph{IP fragmentation}.

Additionally, we cannot determine the origin of failures (see~\autoref{sub:failure-rates}) and high delays along the paths (see~\autoref{sub:response-times}) in more detail, as \ac{RA} probes do not provide such information in the measurement results.
We also acknowledge that \emph{Probe} resolvers with private IPv4 addresses (\eg \acp{CPE}) may use either ISP or public \ac{DNS} services as upstream \ac{DNS}, which we cannot further differentiate with the measured data (see~\autoref{sub:failure-rates}).

As this paper provides a unique insight at \ac{DNS} over TCP, it is also limited to a View from the Edge.
Hence, we plan to extend our study in order to obtain a more complete picture:
First, we intend to measure domains for which we operate the authoritative server, which will allow us to have more fine-grained control regarding measured properties like \emph{EDNS(0) Buffer Sizes} and the actual size of \ac{DNS} responses.
While this will contribute a view of \ac{DoTCP} between resolvers and authoritative servers, we will also explicitly study \emph{truncation} and \emph{IP fragmentation} on stub to recursive resolvers by issuing \ac{DNS} queries to controlled recursive resolvers.
Moreover, this setup will enable us to study the benefits of \emph{\ac{TFO}} and \emph{EDNS0 TCP keepalive}, and their effects on application layer protocols.
Since QUIC, as a reliable, end-to-end encrypted transport protocol, is designed to improve on several shortcomings of TCP, \ac{DoQ} can potentially obsolete the necessity to fall back to \ac{DoTCP} altogether.
We recently presented a study on \ac{DoQ}~\cite{doq_pam}, where we focused on a comparative analysis between \ac{DoQ}, \ac{DoT}, \ac{DoH}, \ac{DoUDP}, as well as \ac{DoTCP}.
However, the study is limited to a single vantage point, which is why we plan to measure \ac{DoQ} from the Edge in order to draw additional comparisons with \ac{DoTCP}. 
This will allow us to provide a more holistic view of \ac{DNS} variations to complement comparable studies on \ac{DoT} and \ac{DoH}.

\section{Conclusion}
\label{sec:conclusion}

We presented a unique view on \acf{DoTCP} from the Edge using 2,500 \acf{RA} probes deployed in residential home networks around the globe.
Based on 12.1M DNS requests issued to \emph{Public} and \emph{Probe} recursive resolvers over \acf{DoUDP} and \ac{DoTCP}, we evaluated \emph{Response Times} as well as \emph{Failure Rates} of \ac{DoTCP}.

We showed that \emph{Response Times} are highly varying depending on the continent of the probes location for \emph{Public} and \emph{Probe} resolvers, and that \ac{DoTCP} is generally slower than \ac{DoUDP}.
Although this was expected, the relative increase in \emph{Response Time} is less than 37\% for most resolvers.
While \ac{TFO} and \emph{EDNS0 TCP keepalive} can be leveraged to further reduce the \ac{DoTCP} response times, we showed that support on \emph{Public} resolvers is still missing, hence leaving room for optimizations in the future.

Analyzing \emph{Failure Rates}, we determined that \emph{Public} resolvers generally have comparable reliability for \ac{DoTCP} and \ac{DoUDP}.
However, \emph{Probe} resolvers show a significantly different behavior, as their failure rate for \ac{DoTCP} is considerably higher with 74.96\%; as a result, \ac{DoTCP} queries targeting \emph{Probe} resolvers fail in 3 out of 4 cases.
Therefore, \emph{Probe} resolvers largely do not comply with the standard described in RFC 7766, which states that all DNS implementations MUST support both \ac{DoUDP} as well as \ac{DoTCP}.
As such, \emph{Probe} resolvers face issues with fallback to \ac{DoTCP} in case of large \ac{DNS} responses to date.
This problem will only aggravate in the future: As \ac{DNS} response sizes will continue to grow, the need for \ac{DoTCP} will solidify.

\section*{Acknowledgements}
We thank Jan Rüth, Alessandro Finamore, Matteo Varvello, and the anonymous reviewers for their insightful feedback and guidance.
In addition, we thank RIPE Atlas for providing the measurement infrastructure.
This work was partly supported by the Volkswagenstiftung Niedersächsisches Vorab (Funding No. ZN3695).

\bibliographystyle{ACM-Reference-Format}
    \bibliography{001_references}

\appendix
\section*{Appendix - Reproducibility}
\label{sec:appendix}

In order to enable the reproduction of our findings, we make the raw data of our measurements as well as the analysis scripts and supplementary files publicly available on GitHub: 
\begin{center}
    \small
    \textbf{\url{https://github.com/kosekmi/2022-ccr-dns-over-tcp-from-the-edge}}
\end{center}

This section gives an overview over the contents of the repository. More details are provided in the \texttt{README.md} within the GitHub repository. \newline

\textbf{Repository Overview}
\begin{itemize}
    \item The file \texttt{analysis.ipynb} is a jupyter notebook containing all analyses detailed in the paper.
    \item The supplementary file \texttt{public-resolvers-ipv4s.csv} is single column text file containing a list of known public resolvers (used in related work).
    \item The supplementary file \texttt{pyasn.dat} is a 2 columns text file mapping RIPEAtlas (RA) probes IP address to the related ASN.
    \item The file \texttt{measurements.parquet} contains the full measurements campaign run via RA probes.  \newline
\end{itemize}

\textbf{Analysis}
\begin{itemize}
    \item Open the Jupyter Notebook \texttt{analysis.ipynb}.
    \item Run the Jupyter Notebook. Depending on machine capabilities, this can take from several minutes up to a few hours for the full dataset.
\end{itemize}

\end{document}